# Surface oxide on thin films of yttrium hydride studied by neutron reflectometry

T. Mongstad[1,*], C. Platzer-Björkman[2], J. P. Mæhlen[1], B. C. Hauback[1], S. Zh. Karazhanov and F. Cousin[3]

[1] Institute for Energy Technology, P.O. Box 40, 2027 Kjeller, Norway

[2] Uppsala University, Solid State Electronics, Box 534, SE-751 21 Uppsala

[3] Laboratoire Léon Brillouin, CEA Saclay, F-91191 Gif sur Yvette, France

Abstract

The applicability of standard methods for compositional analysis is limited for H-containing films. Neutron reflectometry is a powerful, non-destructive method that is especially suitable for these systems due to the large negative scattering length of H. In this work we demonstrate how neutron reflectometry can be used to investigate thin films of yttrium hydride. Neutron reflectometry gives a strong contrast between the film and the surface oxide layer, enabling us to estimate the oxide thickness and oxygen penetration depths. A surface oxide layer of 5-10 nm thickness was found for unprotected yttrium hydride films.

Characterization of the distribution and content of light elements in thin-film systems is difficult using methods based on X-rays, electrons or ion scattering. Hydrogen, the lightest element, is often considered *invisible* due to its low scattering power and the absence of core electrons. Another issue is the high mobility of the H atoms and the reactivity of oxygen to exposed surfaces, resulting in high risks of the sample being modified during sample preparation and/or by the actual measurement. However, H can be the main constituent in solid materials, e. g. metal hydrides. H impurities and H-containing layers are also important for modulating the electrical properties of semiconductors[1]. It is therefore of prime interest to establish methods of investigation for H-containing thin films and their interfaces with other materials.

In this work we show how the use of neutron reflectometry (NR) enables us to investigate the surface oxide and the O and H distribution in films of yttrium hydride. Surface layers are important for the behavior and chemical stability of thin-film metal hydrides. A Pd cap layer is often used for oxidation protection and to catalyze hydrogen uptake. For samples prepared for optical and electrical measurements, the Pd layer is, however, often so thin that the surface partly oxidizes[2]. The thickness of the oxide layer and distribution of the oxygen in the film are hard to assess using conventional methods for compositional analysis as Rutherford backscattering (RBS).

We have in this work investigated unprotected yttrium hydride films by NR. Yttrium hydride appears in two different crystal structures at room temperature and ambient pressure[3]. The two phases have very different electronic states; metallic as $YH_2$ and optically transparent, semiconducting with a band gap of 2.6 eV, when the stoichiometry approaches $YH_3$[4]. Yttrium hydride is highly reactive towards O, but unprotected films of $YH_x$ have proven to be surprisingly stable against oxidation under ambient conditions[5]. The unprotected $YH_x$ films have showed interesting effects in the crystal structure[5] and a strong and technologically relevant photochromic effect has been observed[6].

NR is an efficient way of characterizing thin films, multilayers and interfaces[7]. In particular, it reveals properties of layers and interfaces that are buried below several other layers. NR is a non-destructive technique where the weak sample interaction of the low energy neutrons (~ a few meV) ensures that the sample is not changed by the measurement itself.

NR for characterization of hydrogen in thin films and multilayers of metal hydrides was first reported in 1993 by Mâaza et al[8]. Later, Munter et al. showed how deuterium absorption could be monitored *in-situ* during loading experiments on thin Pd films[9,10]. More recently, a series of studies has been published by Fritzche et al., showing how NR can be used to investigate the thermodynamics of hydrogen uptake process in Pd-capped metal alloy films[11–15]. NR has also proved an efficient way to characterize deviations in the hydrogen stoichiometry in very thin interface layers[9,16], and can be used to determine inter-diffusion of elements at the interfaces of metal hydride films[11,12].

NR relies on the difference in the coherent scattering length density (SLD) for neutrons between different layers in a multilayer structure. The SLD of each layer can be calculated using the formula $\rho = \sum N_i b_i$, summing over all present elements $i$, where $N_i$ is the atomic density of the element and $b_i$ is the coherent scattering length of the element (see Table I). Owing to the negative scattering length density of H, metal hydrides show a large contrast in the SLD with respect to the substrate. O has a positive scattering length density and there is thus also a large difference in the SLD of the hydride film and the surface oxide. The large contrast and large difference in thickness of the layers makes NR ideal for studying these systems. Fig. 1 demonstrates calculated ideal SLD of the materials explored here. For comparison, the calculated scattering length density for X-rays is showed in Fig.1(b). The X-ray SLD is very similar for yttrium oxide and the yttrium hydrides, making the distinction between oxide and hydride difficult in the interpretation of X-ray reflectometry (XRR) data.

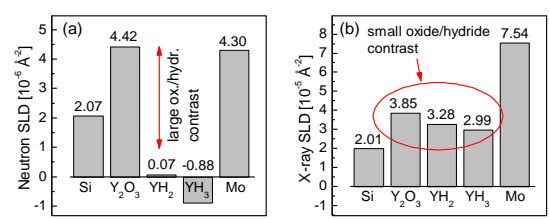

Fig. 1 – (Color online) Calculated neutron and X-ray (Cu Kα) SLD of the materials investigated in this work.

Table I – Coherent scattering lengths for neutrons of the constituent elements[a].

| Element | Scattering length (fm) |
|---|---|
| H | -3.739 |
| O | 5.803 |
| Si | 4.1491 |
| Y | 7.75 |
| Mo | 6.715 |

[a] V. F. Sears, Neutron News 3, 26 (1992).

The NR measurements were carried out at the EROS[17] time-of-flight reflectometer at Laboratoire Leon-Brillouin, Saclay, France. This reflectometer uses a broad wavelength neutron beam with wavelengths from 3 Å to 25 Å. The typical data collection time was 1 hour per sample. The data analysis was done with the EROS reflectivity data analysis software package. The films were deposited on 50mm Ø, 5 mm thick crystalline Si substrates, and the thickness of the investigated films was in the range of 100-150 nm. The samples were deposited by reactive sputtering (see Ref. 5), using 20% $H_2$ as reactive gas in the Ar plasma. Two electronic states of $YH_x$ were obtained depending on the deposition parameters; the conductive and black state (B-$YH_x$) and the optically transparent semiconducting state (T-$YH_x$)[5]. A cap layer of Mo was deposited *in-situ* on some of the samples in order to protect the samples from oxidation in air.

Fig. 2(a) shows the experimental data and the model fit to the NR spectra obtained for a sample of B-$YH_x$. The calculated reflectivity of a pure silicon substrate is also showed in order to illustrate how the film influences the reflectivity spectrum. The critical scattering vector $Q_c$ is not shifted for the sample with respect to the silicon substrate. This is typically observed for a layer with lower SLD than the substrate. A layer of ~100 nm with a larger SLD then the substrate would have caused the appearance of a pseudo-plateau, *i.e.* a shift of the plateau of total reflection towards higher values of the scattering vector $Q$. The interference fringes visible in the spectrum at higher $Q$ values are known as Kiessig fringes[18], which constitute the basis for the fit of the SLD profile (inset in Fig. 2(a)). The bulk of the film is found to have a SLD of $1.1 \times 10^{-6}$ Å$^{-2}$. This is higher than the expected SLD of $YH_2$ (see Fig. 1(a)), and can be explained by incorporation of oxygen in the structure and understoichiometry in the hydrogen content. The fit to the NR data in Fig. 2(a) gives a surface oxide and/or hydroxide layer with a thickness of 5 nm, with a maximum SLD of $2.5 \times 10^{-6}$ Å$^{-2}$. The maximum SLD is limited by the fact that the estimated roughness (1.5 nm on the upper surface) is similar to the thickness of the surface layer. In addition, it is probable that the surface layer contains a considerable amount of H, which will also lower the SLD.

Fig. 2(b) shows data and fit for an equivalent film as in Fig. 2(a), but covered with a 10 nm layer of Mo. The presence of the 10 nm capping layer is clearly visible in the NR data as a large Kiessig fringe superimposed on the smaller oscillations. The upper surface is also found to have a roughness of 1.5 nm, and the maximum SLD in the layer is of $4.0 \times 10^{-6}$ Å$^{-2}$. To obtain the best fit to the data, a gradient has been incorporated in the SLD profile, going from $1.0 \times 10^{-6}$ Å$^{-2}$ at the substrate interface to $1.3 \times 10^{-6}$ Å$^{-2}$ close to the surface oxide. This variation could be a result of the deposition process or O that has penetrated through the Mo layer. It is however difficult to get a fully conclusive fit of NR data on such small and non-abrupt changes in the SLD profile. The possible interpretations of such a gradient might not reflect the physical state of the sample. The SLD profile of this geometry is similar to the sample with a surface oxide (Fig. 1(a)), but the NR data are substantially different. This underlines the strength of NR in differentiating between layers with small differences in thickness and SLD.

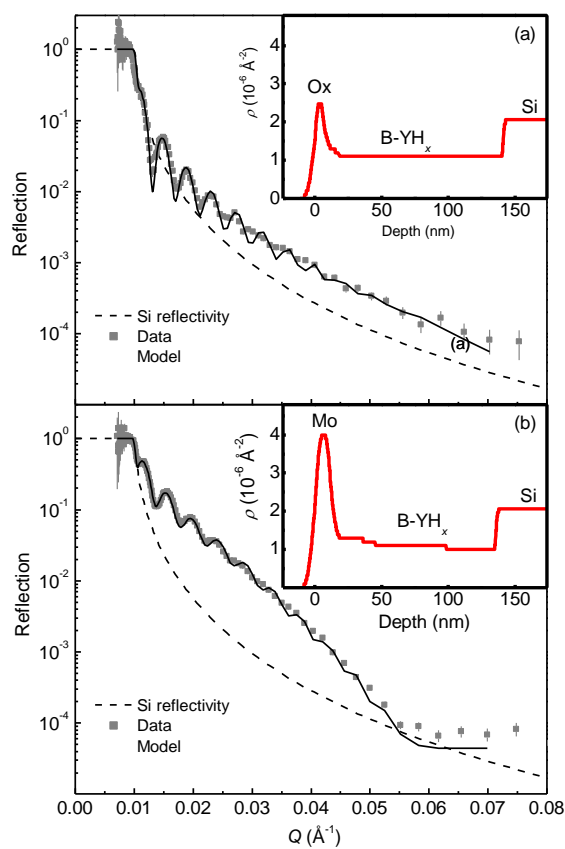

Fig. 2. (Color online) NR data, model and corresponding SLD profile (inset) for (a) an unprotected B-$YH_x$ film and (b) a B-$YH_x$ film covered by 10 nm Mo.

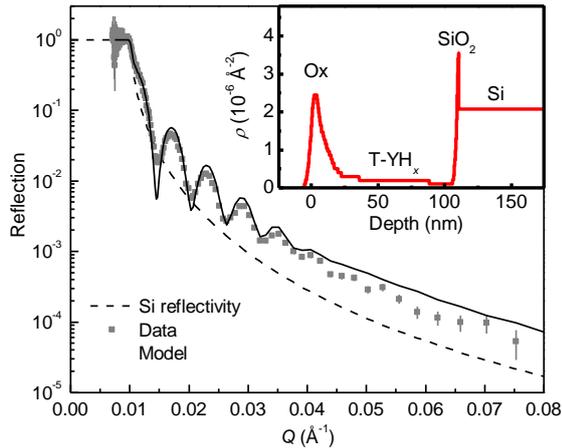

Fig. 3. (Color online) NR data, model and corresponding SLD profile (inset) for T-YH$_x$ film.

Fig. 3 displays data and fit for an unprotected film of T-YH$_x$. The SLD determined for the bulk of the film is ~0.2 × 10$^{-6}$ Å$^{-2}$, which is higher than expected for pure semiconducting YH$_3$ (see Fig. 1(a)). Again this is related to the O content of the film and understoichiometry in H. The surface oxide/hydroxide layer of this sample is around 10 nm, which is higher than for the conductive YH$_x$ film shown in Fig. 2(b). The oxide also has a tail that penetrates into the bulk of the film. This could be related to oxide channels growing at the crystal grain borders that have earlier been discussed for yttrium hydride[2] and pure yttrium[19] films. For pure yttrium films we found a surface oxide layer of ~5 nm and a similar oxygen penetration profile. In Fig. 3, there is a layer of SiO$_2$ with higher SLD present on the substrate interface. The substrate oxide layer is a result of the synthesis of the samples. Because of the poor adhesion of T-YH$_x$ to pure Si, the native surface oxide on the Si substrate was not removed prior to deposition.

Because of the strong sensitivity to H, NR can be used to estimate the H content in a film. The value of the SLD is significantly higher for the B-YH$_x$ (1.1× 10$^{-6}$ Å$^{-2}$, Fig. 2) than for the T-YH$_x$ (0.2 × 10$^{-6}$ Å$^{-2}$, Fig. 3). As the samples are prepared in an approximately identical process and has basically the same crystal structure[5] this provides an unambiguous proof of higher H content in the latter. In principle the H concentration can be directly calculated from the SLD and the crystal structure. However, a true estimation of the H content requires knowledge of the impurity content, porosity and concentration of interstitials and vacancies in the sample. For a trustworthy estimation of the H content, NR therefore needs to be combined with complementary techniques as X-ray diffraction, XRR and RBS. Alternatively, one can take advantage of the isotope contrast between hydrogen and deuterium. D has a positive coherent scattering length for neutrons, and thereby contributes to an increase in the SLD as opposed to H. If two sets of samples are prepared in the same way but one with H the other with D, the difference in SLD can give an accurate estimation of the H/D content of the sample. Similarly, NR has been used to study loading/unloading of H in metallic films, where the SLD prior to and post-loading is compared in order to estimate the H concentration[9,11,13,15,20,21].

In conclusion, NR is highly suitable for the study of oxide layer formation on thin films of metal hydrides, because of the large contrast in SLD between oxide and hydride. NR can similarly also be used for example to investigate H accumulation at interfaces in semiconductors or to study surface oxides on non-hydride films with low SLD.

This work has been made possible through funding from the Norwegian Research Council through the NANOMAT program. The NR measurements were supported by the European Commision under the 7th Framework Programme: Integrated Infrastructure Initiative for Neutron Scattering and Muon Spectroscopy: NMI3/FP7 – Contract No 226507.